\documentclass{article}
\usepackage[utf8]{inputenc}
\usepackage{amsmath,amsfonts,amssymb,eucal,graphicx,graphics,bm,mleftright}
\usepackage[sort&compress,numbers]{natbib}
\usepackage{xparse}
\usepackage{epsfig}
\usepackage{wrapfig}
\usepackage{subfigure}
\usepackage{amssymb}
\usepackage{epsfig}
\usepackage{subfigure}
\usepackage{graphicx,color}
\usepackage{textcomp}
\usepackage{url}
\usepackage{float}
\usepackage{bm}

\title{Metabolic scaling in small life forms }
\author{Mark E. Ritchie \& Christopher P. Kempes}
\date{\today}

\begin{document}

\maketitle

\begin{abstract}
Metabolic scaling is one of the most important patterns in biology. Theory explaining the 3/4-power size-scaling of biological metabolic rate does not predict the non-linear scaling observed for smaller life forms. Here we present a new model for cells $<10^{-8}$ m$^{3}$ that maximizes power from the reaction-displacement dynamics of enzyme-catalyzed reactions. Maximum metabolic rate is achieved through an allocation of cell volume to optimize a ratio of reaction velocity to molecular movement. Small cells $< 10^{-17}$ m$^{3}$  generate power under diffusion by diluting enzyme concentration as cell volume increases. Larger cells require bulk flow of cytoplasm generated by molecular motors. These outcomes predict curves with literature-reported parameters that match the observed scaling of metabolic rates for unicells, and predicts the volume at which Prokaryotes transition to Eukaryotes.  We thus reveal multiple size-dependent physical constraints for microbes in a model that extends prior work to provide a parsimonious hypothesis for how metabolism scales across small life.

\end{abstract}

\section*{Introduction}

Understanding how and why organisms differ in their demand for and use of resources is a key objective of biologists \cite{brown2004,enquist2003,kaspari2004} and critical to understanding the response of biodiversity and ecosystem function to global changes  \cite{ritchie2009,pawar2012}. A fundamental and often-debated pattern is how metabolic rate, $B$, scales with body mass, $M$ \cite{brown2004,agutter2011,glazier2010}, followng the form 
\begin{equation}
B=B_0 M^\beta.
\end{equation}

Data for vertebrates and vascular plants \cite{schmidt1984,peters1986,west1997general,west1999general,mori2010mixed}show an average interspecific scaling exponent of 3/4. This relationship has inspired diverse theories \cite{west1997general,west1999general,banavar2002supply,banavar2010,maino2014}, including the network model, which derives 3/4 from the need for organisms to supply the entire body volume with resources from vascular resource distribution networks that minimize energy dissipation. More recent analyses \cite{delong,brown2004,glazier2014,kolokotrones2010,west2005,makarieva2008} that include organisms from the smallest 11 orders of magnitude in size that largely lack vascular distribution networks, show a variable metabolic scaling exponent that changes across size ranges from scaling exponents as high as 2 -- super-linear scaling -- for the smallest range and near 3/4 for the largest organisms. Understanding the basis for this variation is important, as many different organism features can be derived from this metabolic scaling, including how maximum growth rates and ribosomal abundances scale with cell size along with key tradeoffs between features \cite{kempes2012growth,kempes2016evolutionary}. However, there is no general, parsimonious theory derived from physical and chemical principles that addresses this size-dependent scaling for these smallest organisms \cite{delong}.

Here we propose a combined thermodynamic and reaction description of metabolic rate for a cell. We assume that natural selection will favor organisms that can perform more work per time (power) for material synthesis, replication, repair, locomotion, and other cellular functions  \cite{kempes2019scales}. While often co-limited by materials in the environment, optimizing this power faces cellular trade-offs concerning both environmental and intra-cellular physical constraints. Consequently, we optimize cellular features to maximize the free energy produced by the conversion of chemical substrates to products, where heat and products may inhibit metabolism if not moved away from reaction sites. This thermodynamic approach provides a framework for considering the two processes, conversion and displacement, simultaneously. 

\section*{Thermodynamics of reaction and molecular displacement}
We assume that work is proportional to the volume of reactive surface, where macromolecular catalytic enzymes are freely dispersed in the cytoplasm or attached to the membranes, cytoskeleton, and/or other organelles within the entity \cite{demetrius2003}. However, a portion of cell volume is needed to allow substrates from the environment to reach reaction surfaces and to allow displacement of reaction products away from reaction structures. This sets up a conflict between the volume devoted to metabolic processes and that devoted to transport. Here we derive a physical and chemical principles-based theory for metabolic scaling for metabolic rate across the roughly 11 orders of magnitude in volume of organisms that lack branching vascular systems. We describe metabolism from a reaction-displacement thermodynamic system centered on or near reaction surfaces and its generation of free energy within a spherical space otherwise obstructed by surfaces at which reactions occur. Our framework applies to unicellular Prokaryotes, Archaea, and Eukaryotes along with eukaryotic organelles.


Our basic model, which we expand on in stages, is to consider metabolism in a sphere described by the following reaction-diffusion model
\begin{eqnarray}
\frac{\partial A}{\partial t}&=&D\frac{1}{r^{2}}\frac{\partial}{\partial r}\left(r^{2}\frac{\partial A}{\partial r}\right)-kA Z \nonumber \\
\frac{\partial P}{\partial t}&=&D\frac{1}{r^{2}}\frac{\partial}{\partial r}\left(r^{2}\frac{\partial P}{\partial r}\right)+kA Z 
\label{spherical-pde}
\end{eqnarray}
where the product, $P\left(r\right)$, substrate, $A\left(r\right)$, and enzyme, $Z\left(r\right)$, concentrations (mol m$^{-3}$) are all a function of radius, $r$, inside a sphere, and where the dynamics are one dimensional in spherical coordinates with spherical symmetry.

Additionally, $k=k_{cat}/K_{M}$ is the reaction constant ((m$^{3}$ mol $^{-1}$ s$^{-1}$)) and $D$ is a displacement coefficient (e.g. molecular diffusivity in some cases) and has units of (m$^{2}$ s$^{-1}$) so that each rate is a change in concentration (mol m$^{-3}$ s$^{-1}$). We consider the steady-state dynamics, representing a persistent entity and its metabolism in a fixed environment over time, allowing us to solve for closed-form solutions of $P(r)$ and $A(r)$ under a given concentration of $Z$ (see SI). The steady-state free energy production at any location in the cell is given by
\begin{equation}
J^{*}\left(r\right)=  k R T A^{*} Z \ln\left(\frac{A^{*}K_{eq}}{P^{*}}\right)
\end{equation}
where $K_{eq}$ is the equilibrium constant of the energy producing reaction (which can be interrelated with $\Delta G_{0}$), $T$ is temperature, and $R$ is the ideal gas constant. The entire metabolism (given as a power in watts) of the entire cell is then described by 
\begin{equation}
B= 4 \pi k R T Z \left(1-\frac{3 V_{e}(r_{c})}{4\pi  r_{c}^3}\right)
    \int_0^{r_{c}} x^2 A(x) \ln \left(\frac{K_{eq} A(x)}{P(x)}\right) dx
\end{equation}
where $r_{c}$ is the radius of the cell and $V_{e}$ is the essential volume required for other for other cellular materials, such as DNA, and is unavailable for energy generation.

In this system, we can maximize power (free energy/time), $B$ (Watts), as a function of $Z$. If there were no tradeoffs, then $B$ would be maximized by the largest feasible $Z$ at a given size (a cell full of enzymes). However, the need to move substrate into different regions of the cell and to displace products away from reaction sites introduces a trade-off between the effective diffusion coefficient, $D$, and enzyme concentration, $Z$. An increase in $Z$ corresponds to a decrease in diffusivity \cite{banks2005anomalous,roosen2011protein}, and this leads to an optimal enzyme concentration corresponding to a maximal metabolic rate (see SI).

\section{Optimal tradeoffs}
\begin{figure}[h!]
    \centering
    \includegraphics[width=0.49\textwidth]{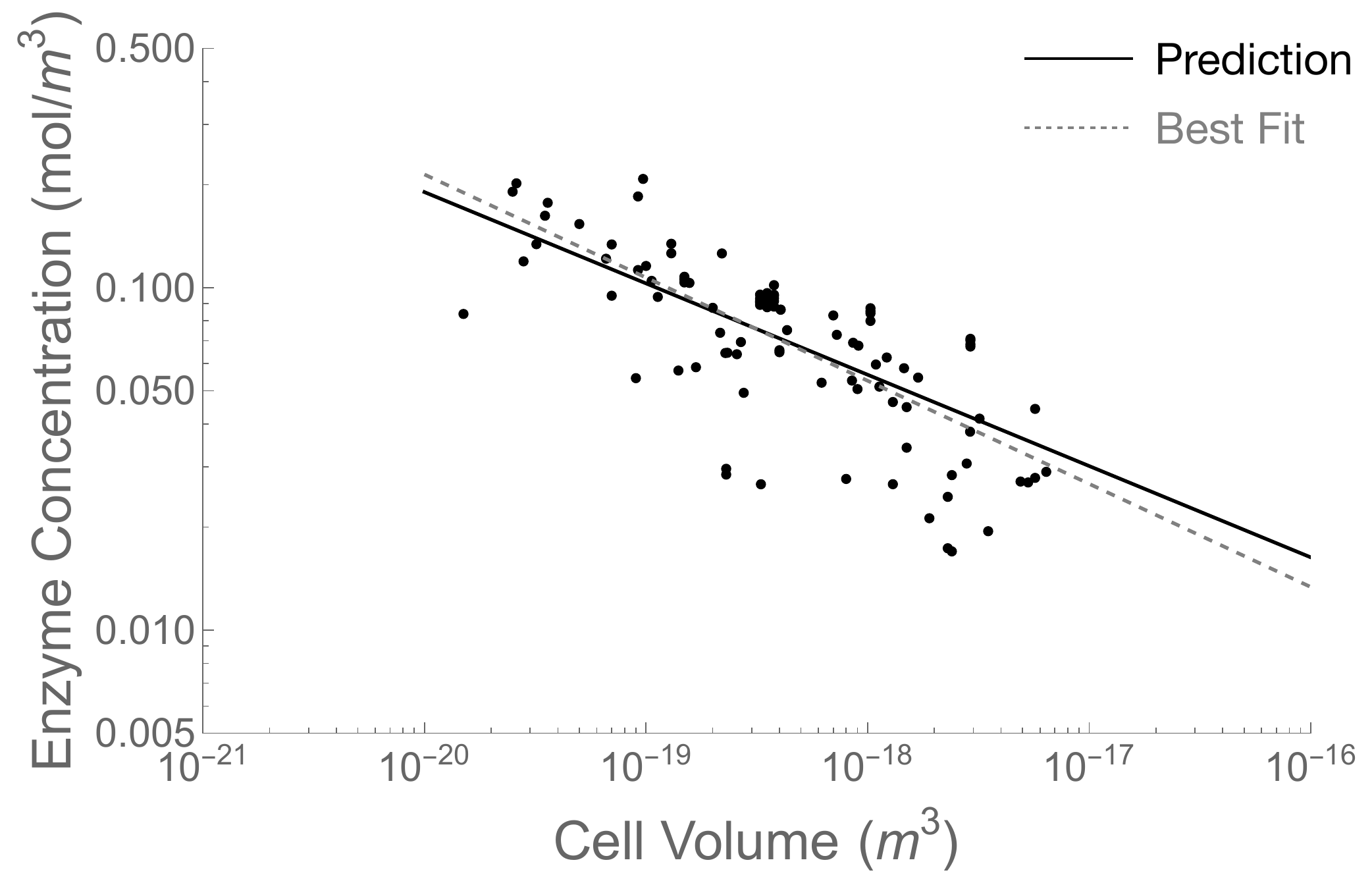}
    \includegraphics[width=0.49\textwidth]{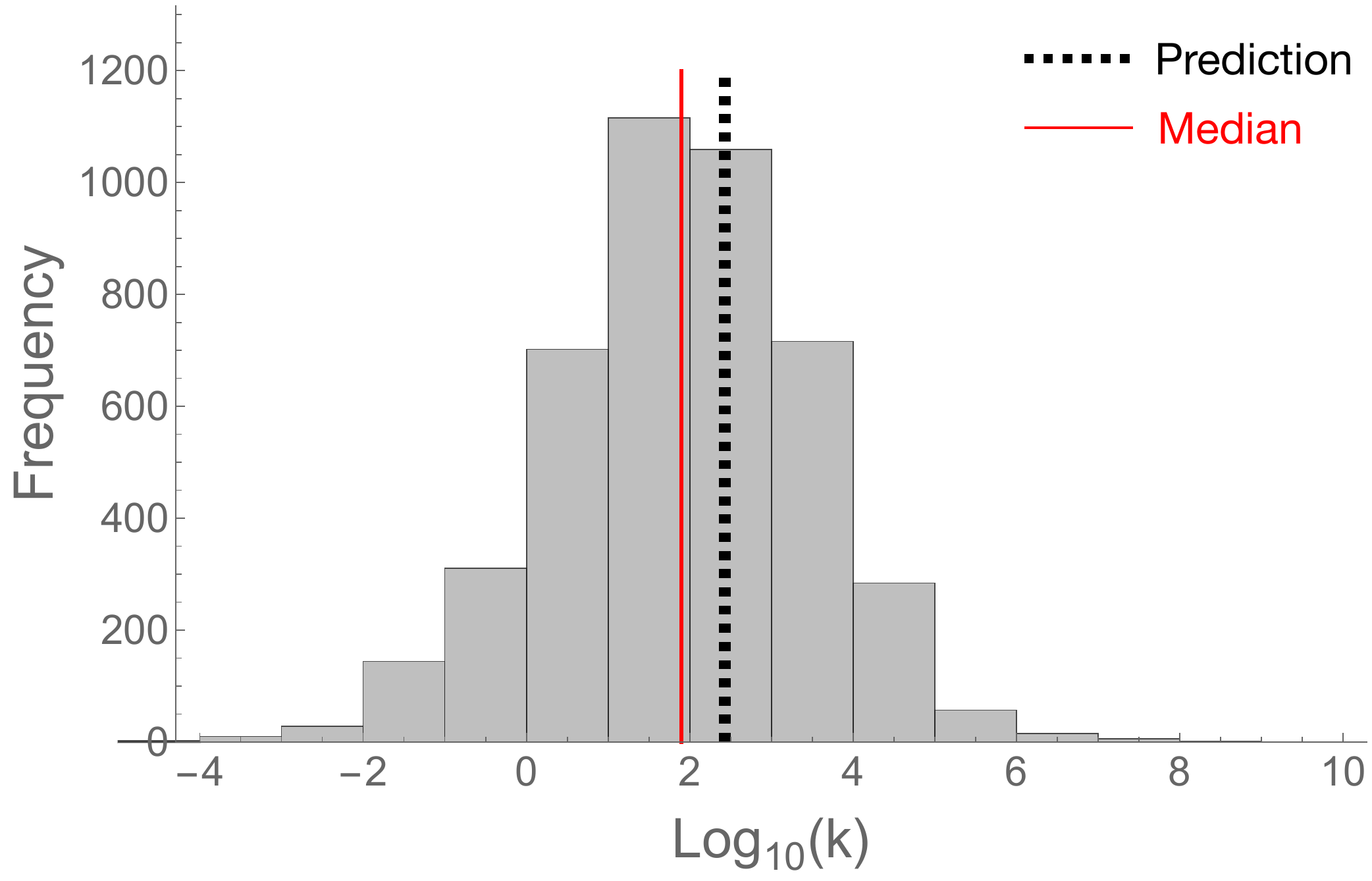}
    \caption{(A) The scaling of the concentration of a single enzyme with cell volume. The solid line is the prediction from our theory (exponent $= 0.27$) and the dashed line is the best fit to data (exponent $= 0.30 \pm 0.06$) compiled across the range of bacteria (calculated from total protein abundance scaling in \cite{kempes2016evolutionary}). (B) The distribution of $k$ values for a few thousand enzymes \cite{bar2011moderately} with the median value show in red compared to the best fit of our model (dashed black line).}
    \label{Fig:concentration}
\end{figure}

Maximizing power over the whole cell, at a given cell size $r_{c}$ subject to variation in the concentration of enzymes, $Z$, we find that the optimal enzyme concentration $Z^{*}$ follows\begin{equation}
Z=\frac{1}{e}\left(\frac{6 K_{eq} D_{0}}{k r_{c}^{2}}\right)^{1/\left(1-\gamma\right)}
\label{EQ:enzyme-scaling}
\end{equation}
where $\gamma$ is the scaling between diffusivity and enzyme concentration at high enzyme concentrations, and $D_{0}$ is the normalization constant for that scaling relationship (see SI). Remarkably, this relationship shows that the optimal enzyme concentration in cells can be predicted from a few fundamental thermodynamic, $K_{eq}$, and kinetic, $k$, constants (Figure \ref{Fig:concentration}). Note that $Z^{*}$ depends on a ratio of reaction rate, $k$, to a diffusivity normalization for small molecules in the cytoplasm, $D_{0}$.

Using the known relationship for diffusivity with enzyme concentration, where $\gamma \approx -3/2$ for sufficiently high concentrations \cite{roosen2011protein}, we obtain
\begin{equation}
Z = Z_{0} V_{c}^{-4/15}.
\end{equation}

Previous studies have shown that total protein count scales with cell size following a power law with an exponent $< 1$ \cite{kempes2016evolutionary}, but these relationships have no fundamental explanation. Our optimization predicts that protein concentration should become more dilute as cells become larger following an exponent of $-4/15\approx -0.27$, which is indistinguishable from the best fit exponent to data of $-0.30\pm0.06$ (Figure \ref{Fig:concentration}).

The prediction for $Z$ was determined by optimizing metabolic rate and so we also predict the relationship for maximum metabolic rate:
\begin{equation}
B = B_{0} V_{c}^{11/15} \left(1-\frac{v_{e}}{V_{c}}\right).
\label{EQ:ProkBscaling}
\end{equation} 
Here $B_{0}$ is a constant that is a complicated function of diffusivity, kinetic and thermodynamic parameters (see SI), and $V_{e}$ is the volume of other essential macromolecules which follows $V_{e}= v_{0} V_{c}^{\alpha}$  with $\alpha=0.83$ \cite{kempes2016evolutionary}.  

\begin{figure}[h!]
    \centering
    \includegraphics[width=0.75\textwidth]{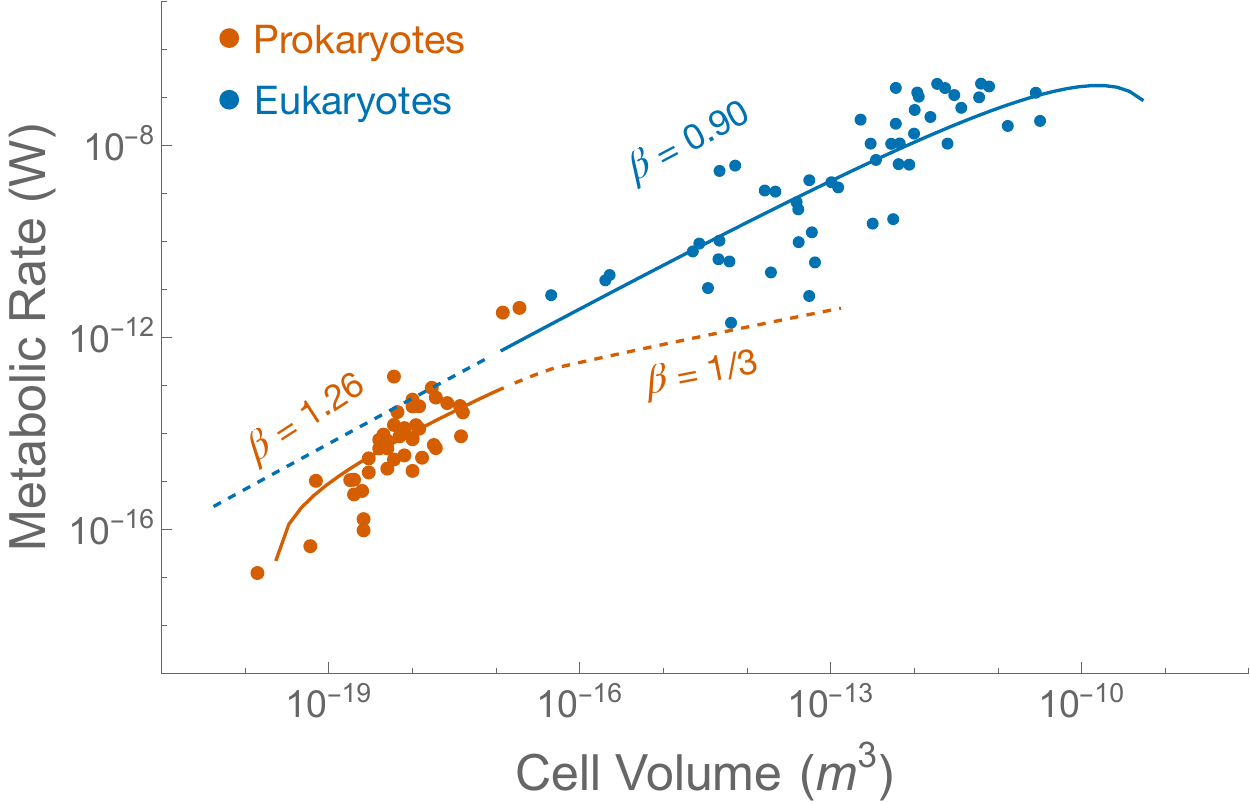}
    \caption{The prediction for prokayote (red) and eukaryote (blue) metabolic rate along with compiled data \cite{delong}. The red dashed curve to the right of the Prokaryote data indicates the scaling of metabolism with an exponent of $1/3$ once diffusivity has saturated to the molecular maximum value. Likewise curvature of the bulk flow model at the largest sizes results from dilution of enzymes by increasing densities of molecular motors.}
    \label{Fig:scaling}
\end{figure}

\section*{Metabolic scaling}

Equation \ref{EQ:ProkBscaling} provides excellent agreement with observed metabolic rates for prokaryotes (Fig. \ref{Fig:scaling}). Metabolic rate has strong, non-power law curvature for small cell sizes driven by the volume $V{e}$ required for macromolecules not involved in reactions, such as DNA, and this is what leads to the apparent super-linear ($\beta > 1$) scaling previously observed for prokaryotes \cite{delong}. This curvature quickly relaxes for larger cells due to the sublinear ($\beta < 1$) scaling of $V_{e}$. For cell sizes up to about the volume of {\it E. coli}  ($\approx10^{-18}$ m$^{3}$) the power law approximation is superlinear (SI Figure \ref{SIFig:aproximate-exponent}) with an approximate exponent of $1.68$ which agrees well with the previous estimates of $1.75$ \cite{delong}.

For larger bacteria, the scaling of $B$ converges on the sublinear scaling of $11/15 = 0.73$. For intermediate prokaryote cell sizes, binning the data to remove oversampling of certain cell volumes and scatter, shows exceptionally good agreement between our model and the data (SI Figure \ref{SIFig:binned-comparison}). These results are not in conflict with the overall superlinear fit to all of the  prokaryotic data \cite{delong}, but show that this is a consequence of a complicated set of scale transitions and shifting constraints. Thus, downstream models that derive results from the scaling exponent of metabolic rate (e.g. \cite{kempes2012growth,kempes2016evolutionary}) will gain accuracy and predictive power by starting from our more complex model. For example, models of growth which assume a power law for metabolism \cite{kempes2012growth} should incorporate the curvature described in Equation \ref{EQ:ProkBscaling}.

\section*{Size-dependent failure of diffusion and emergence of transport mechanisms}
The trade-off between decreasing enzyme concentration and increased diffusivity applies only for high concentrations of the enzyme. Eventually, for large enough cells, sufficiently low enzyme concentrations are reached such that $D=D_{max}$, the maximum diffusivity of small molecules in water (see SI). Once diffusivity saturates, the optimal concentration follows $Z\propto V_{c}^{-2/3}$ (see SI) yielding $B\propto V_{c}^{1/3}$ (Fig. \ref{Fig:scaling}, dashed red curve) and a metabolic rate controlled by diffusion across the cell membrane. Such scaling also avoids thermodynamic shutdown (power = 0) that can occur in the center of larger cells. Shutdown can occur when $D_{max}$ is not fast enough to sufficiently displace reaction products and avoid reaction reversal. These changes may result in body designs, such as large central vacuoles, filamented rods, membrane bound organelles, and extreme polyploidy \cite{schulz1999dense,lane2010energetics,kempes2016evolutionary,volland2022centimeter} that minimize the potential for interior thermodynamic shutdown, but yield a relatively slow metabolic rate for prokaryotes larger than $10^{-17}$ m$^{3}$ \cite{kempes2016evolutionary}. 

With further increases in cell volume, metabolic rate can only be increased by increasing the rate of molecular displacement, as $Z$ is well below the concentration that influences diffusivity.  A variety of means for increasing transport in cells exist (e.g. \cite{luby1999cytoarchitecture,hurtley1998cell,xu2019direct,fakhri2014high,brangwynne2009intracellular,bressloff2013stochastic,mogre2020getting,qu2021persistent,bowick2022symmetry,guo2014probing}) with most attributable to the addition of structures, such as molecular motors or transport proteins, that each generate active transport of molecules or viscous bulk flow of cytoplasm over at least some local region within the cell. We model this as an enhanced effective diffusivity of $D_{trans}$. For example, the movement of molecules along a cytoskeletal network and randomly arranged regions of bulk flow with a random flow direction can each be thought of as random walks with long single steps, and so displacement in the simplest summary model for all of these processes can be described as an enhanced diffusion (e.g. \cite{luby1999cytoarchitecture,lipowsky2001random,pierobon2006bottleneck,brangwynne2009intracellular,bressloff2013stochastic,fakhri2014high,bowick2022symmetry,guo2014probing}). A new trade-off is introduced between the space occupied by dedicated transport structures versus the space devoted to catalyzing reactions (enzymes and supporting structures). This effective diffusivity and volume tradeoffs are captured by
\begin{equation}
D^{\prime}=(1-\rho)D_{max}+\rho D_{trans}
\end{equation}
and
\begin{eqnarray}
\frac{\partial A}{\partial t}&=&D^{\prime}\frac{1}{r^{2}}\frac{\partial}{\partial r}\left(r^{2}\frac{\partial A}{\partial r}\right)-k_{cat}A Z  \nonumber \\
\frac{\partial P}{\partial t}&=&D^{\prime}\frac{1}{r^{2}}\frac{\partial}{\partial r}\left(r^{2}\frac{\partial P}{\partial r}\right)+k_{cat}A Z_0 
\label{euk-pde}
\end{eqnarray}
where $D_{trans}$ is the enhanced diffusivity in the region affected by the active transport or bulk flow, $\rho$ is the fraction of the average cellular volume that is associated with active transport or bulk flow, and $\epsilon$ is the ratio of the effective volume of active transport to the volume of the structures generating that transport (e.g. the ratio of the volume of a region of enhanced transport to a molecular motor or transport protein's volume). 

We optimize total metabolic rate (see SI) at a given cell volume as a function of $\rho$ and find that $\rho$ grows with cell size following a complicated function (see SI) that scales like $\approx V_{c}^{0.10}$ for small cell sizes and $\approx V_{c}^{2/3}$ in the limit of large cell sizes. This density scaling implies that the volume dedicated to transport structures scales super-linearly with overall cell size ($\approx V_{c}^{1.1}$  to $\approx V_{c}^{5/3}$).Thus, for cell sizes larger than $10^{-17}$ m$^{3}$, the effective diffusivity parameter can increase with cell volume but at the expense of an ever-increasing proportion of cell volume devoted to transport. This yields a size-dependent curve for maximum metabolic power in logarithmic space (blue curve in Fig. 3) that fits the available data for metabolic rate of Eukaryotes (again with average transport and reaction kinetic parameters). Over the middle size range of the single cell Eukaryotes, metabolic rate should approximate a power law with an exponent of $0.90$ which agrees very well with the power law fit to data of $0.90 \pm 0.17$ \cite{delong,kempes2012growth}. However, the super-linear scaling of the required volume of transport molecules means that motors and transport proteins begin to substantially reduce the volume for reaction structures. This transport volume requirement ultimately may limit metabolic rate, and impose a theoretical upper limit to Eukaryote size of $10^{-9}$ m$^{3}$ which compares well to the size of the largest unicellular organisms in the metabolic database (Fig. \ref{Fig:scaling}).

\begin{figure}[h!]
    \centering
    \includegraphics[width=0.85\textwidth]{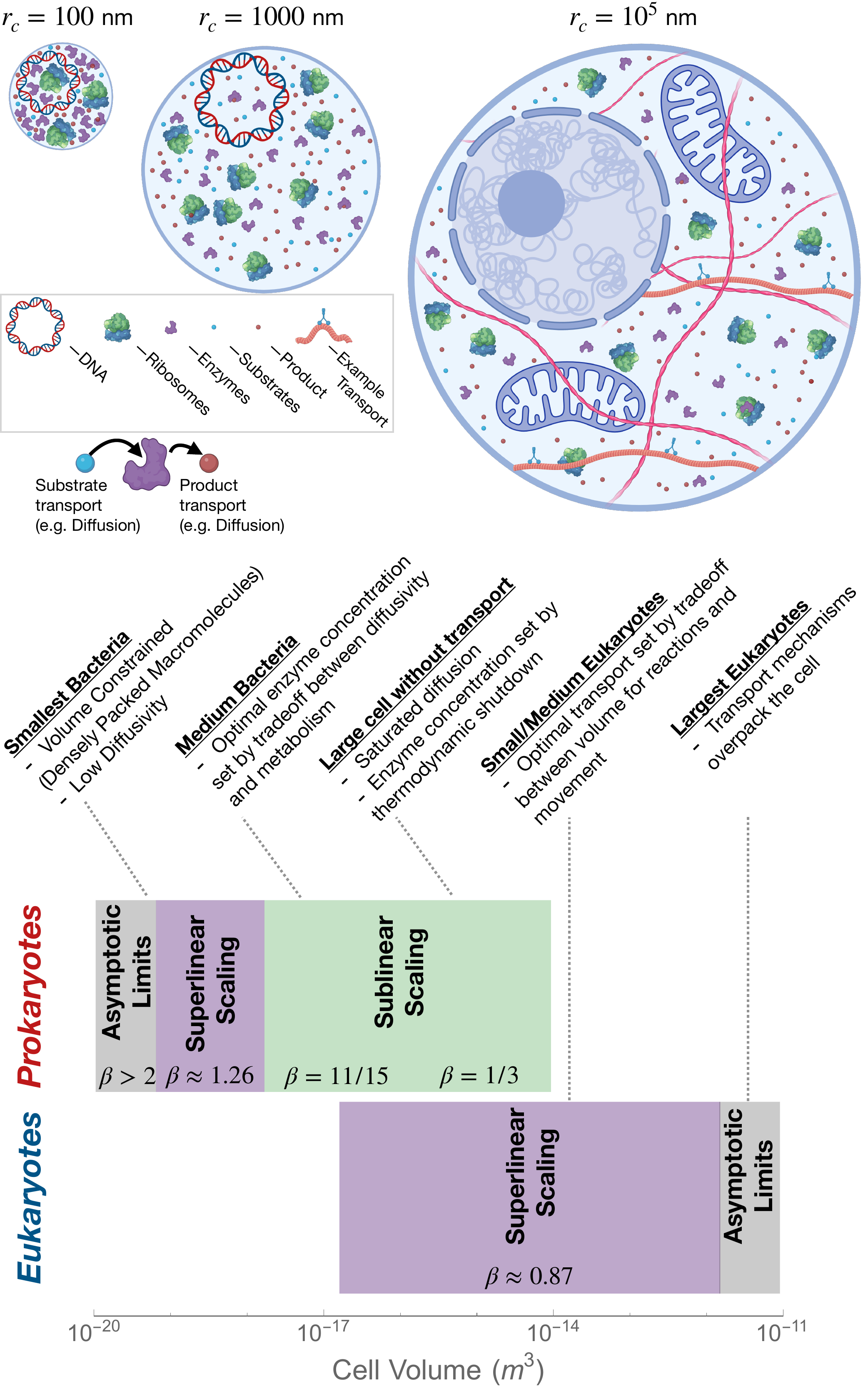}
    \caption{Representations of cells with increasing cell radius filled by cell components DNA, enzymes, ribosomes, and eventually molecular motors \cite{biorender}. The scaling regimes and key limitations faced by cells of each category are also indicated.}
    \label{Fig:overview}
\end{figure}
 
\section*{Discussion}

These analyses produce a series of metabolic scaling curves (Fig. \ref{Fig:scaling}) across a succession of cell (or organism) volume ranges. Each curve is associated with a particular mode of reaction volume and molecular displacement mechanisms, with molecular diffusion at the smallest sizes then with an enhanced diffusion within devoted transport regions at the largest sizes (Fig. \ref{Fig:overview}).

Across the size spectrum, optimal metabolic rate is set by the tradeoff between volume for transport and volume for metabolic reactions. For prokaryotes, the smallest cells require space for other macromolecules such as DNA, which limit the space available for catalytic reaction enzymes. The largest unstructured prokaryotes face the slow scaling of metabolism set by the maximum diffusivity of small molecules in water. For unicellular eukaryotes, transport molecules provide the needed increase in molecular movement and metabolism scales almost linearly with cell volume. At the largest sizes, unicellular eukaryotes reach a sharp decrease dictated by the required overpacking of transport structures to assure the necessary effective diffusivity throughout the cell. Collectively, these curves merge to form an overall nonlinear relationship between metabolic rate and size that fit observed data very strongly (Fig. \ref{Fig:scaling}) and predict the scaling of protein and other material concentrations and the size at which major transitions in body plans occur. 

These curves clearly depart from the prediction of a single scaling exponent of $3/4$ predicted by the theory for the optimal organization of vascular networks, and support the general hypothesis that the smallest forms of life face different physical constraints, with different accompanying solutions, for maximizing metabolic power. However, the derived mechanisms here tend, as a first approximation, to a sublinear scaling not far from $3/4$. For example, the mid range bacteria are predicted to scale like $11/15=0.73$, and the largest eukaryotes scale like $\approx .80$ as curvature sets in. These scalings for both small life and for organisms with vascular systems occur for very different mechanistic reasons, but are linked by the fundamental need for life to allocate and organize space to achieve sufficiently high rates of molecular displacement. Our results thus support more general arguments for why metabolic scaling should not be either surface area to volume scaling ($\beta = 2/3$) or strictly proportional to volume (isometric, $\beta = 1$).  

Our work suggests that previous analyses of prokaryote metabolic scaling with all of the data \cite{delong}, which found super-linear scaling, requires size-dependent shifts in physical constraints that accompany shifts in scaling to approximate a super-linear power law. Our results demonstrate that the overall super-linear pattern is driven by strong curvature away from a power law at the smallest cell sizes.

Downstream extension of the fits using all data have been used to derive other features such as the increase in growth rate with cell size and the total abundance of ribosomes \cite{kempes2012growth,kempes2016evolutionary}. Such models may gain accuracy and improved resolution in size ranges by incorporating the more complicated metabolic scaling derived here. In addition, other cellular constraint perspectives have been proposed for explaining scaling in bacteria such as biosynthetic costs of the membrane \cite{trickovic2022resource}, ribosome and protein abundances and costs \cite{lynch2015bioenergetic,kempes2016evolutionary}, the spatial location and number of organelles and genomes \cite{delong,lane2010energetics}, the increasing number of genes and their cost \cite{delong,lane2010energetics,lynch2015bioenergetic,alberti2017phase}, and transporter optimizations associated with the environment (e.g. \cite{kempes2019scales} for a review). In addition, the deployment of phase separation as another means for enhancing reaction rates  \cite{azaldegui2021emergence,zwicker2022intertwined} and the spatial architecture of macromolecules \cite{castellana2016spatial,wingreen2015physics} may also be important for cellular scaling. These are all important biophysical, physiological, and evolutionary considerations, and it will be important to integrate these additional constraints with cellular scaling in a more complicated optimization approach \cite{kempes2019scales} that combines these costs with the diffusive tradeoffs employed here. 

Large prokaryotes also deploy active molecular transport and so the two models presented here represent bounding cases where large prokaryotes ({\it E. coli} and bigger) and small unicellular eukaryotes. Multicellular organisms overcome the limit faced by the largest single cells by evolving mechanisms of bulk flow among cells. At its extreme, optimizing such bulk flow through tubular vascular systems to minimize energy dissipation (friction) produces the well-known $B \propto M^{3/4}$ \cite{west1997general,west1999general,west1999fourth,banavar2002supply,savage2010hydraulic,brummer2017general}. Taken together, our results and vascular network models show that, organisms across the tree of life may face universal tradeoffs between biochemical reactions and the space devoted to transport, where larger organisms must employ increasingly energy-requiring methods to increase molecular transport velocities.

Our results also predict the scaling of materials within the cell. For small entities relying on diffusion, we predict the total mass of enzymes (protein) to scale as $V_{c}^{11/15}$, which closely agrees with observed scaling proportional to $V_{c}^{0.70}$ \cite{kempes2016evolutionary}. We also predict the minimum size and volume scaling at which cells should produce molecular motors, microtubules and other cytoskeletal material and other mechanisms to enhance diffusivity (Fig. \ref{Fig:scaling}). These scaling predictions cannot yet be tested, as they predate the measurement of such volumes or masses for diverse species and cell sizes, with just recently-developed microscopy techniques. However, the largest prokaryotes are known for unusual storage capacity \cite{schulz1999dense,kempes2016evolutionary}, many copies of the genome \cite{lane2010energetics}, and surprising amount of cytoskeleton \cite{rodrigues2023actin}. All of these characteristics may give hints to the eukaryotic transition, and indeed Asgard archaea, which are suggested to be the eukaryotic ancestor, have several atypical cellular morphological characteristics \cite{rodrigues2023actin,avci2022spatial,imachi2020isolation}.

Our results suggest that active transport in and/or bulk flow of the cytoplasm is likely required to generate free energy even in single cell organisms such as larger Prokaryotes and single-cell Eukaryotes. Extrapolation of the required molecular movement in our models further suggests that smaller multicellular organisms may require mechanisms such as contraction of tissues and/or locomotion to generate bulk transport of intercellular fluids without an organized single source vascular system. Extending further, branching pressurized circulatory networks may be required to sustain metabolism for organisms $>1$ g, a hypothesis supported by evidence that smallest mechanical pumping circulatory systems are limited by damping of pulsatile flow \cite{west1997general}. 

Our results now provide a physical and chemical explanation for the unique metabolic scaling exhibited by the smallest $11$ orders of magnitude in organism size. We demonstrate that these life forms face significant design constraints, set by limited space for macromolecules and diffusive molecular movement, and the continuing need for faster molecular displacement mechanisms for the largest unicellulars. The predicted scaling relationships as organisms shift across the different transport limitation domains suggest that natural selection has addressed this sequence of problems during the evolution of larger cell body sizes. Such evolution has led to cells with clustered reaction surfaces, channels of cytoplasm for transporting molecules, molecular motors and cytoskeletal structures to enhance diffusion and then finally by mechanical pumping mechanisms that preview the evolution of powered branched circulatory systems. Our work now extends prior theory to provide a parsimonious hypothesis for how metabolism and the materials that support such activity scale across all life.


\section{Supplementary Material}


\newpage

\section{Diffusions Tradeoffs} 
The coupled diffusion equations to be solved for steady state are 
\begin{eqnarray}
\left(\frac{D(Z) \left(x^2 A''(x)+2 x
   A'(x)\right)}{x^2}-k Z A(x)\right) &=& 0 \\
 \left(\frac{D(Z) \left(x^2
   P''(x)+2 x P'(x)\right)}{x^2}+k Z A(x)\right) &=& 0 
   \end{eqnarray}
with the boundary conditions
\begin{eqnarray}
A(r_{c})&=&A_{0} \\
A^{\prime}(0) &=& 0 \\
P(r_{c})&=&P_{0}.
\end{eqnarray}
The solution to these equations is
\begin{equation}
A\left(r\right)= \frac{A_{0} r_{c} \text{csch} \left(\frac{r_{c}\sqrt{kZ} }{\sqrt{D(Z)}}\right) \sinh \left(\frac{r \sqrt{kZ}}{\sqrt{D(Z)}}\right)}{r}
\label{Ar-solution}
\end{equation}
Following a similar derivation, the corresponding equation for $P$ is
\begin{equation}
P\left(r\right)=A_{0}+P_{0}-\frac{A_{0} r_{c} \text{csch}\left(\frac{r_{c} \sqrt{k Z}}{\sqrt{D(Z)}}\right) \sinh \left(\frac{ r \sqrt{kZ}}{\sqrt{D(Z)}}\right)}{r}
\label{Pr-solution}
\end{equation}

Returning to the integral which represents the full metabolic rate of the cell, can can use these solutions for $A$ and $P$ set constraints on $Z$. To avoid thermodynamic shutdown we need $\frac{K_{eq} A(r)}{P(r)}>1$ which, given the solutions above, leads to the condition
\begin{equation}
r\;\text{csch}\left(\frac{r \sqrt{k Z}}{\sqrt{D(Z)}}\right)<\frac{A_{0} \left(1+K_{eq}\right)
   r_{c}\;\text{csch}\left(\frac{r_{c} \sqrt{k Z}}{\sqrt{D(Z)}}\right)}{A_{0}+P_{0}}.
\end{equation}
Thermodynamic shutdown is a problem for the interior of the cell and so we are interested in checking this condition for $r=0$. Evaluating the above inequality at $r=0$ and expanding in small $r_{c}$ leads to
\begin{equation}
\frac{Z} {D\left(Z\right)} <\frac{6 (A_{0}K_{eq} - P_{0})}{A_{0} k r_{c}^2(1+K_{eq})}.
\end{equation}
This equation isn't fully solved because the diffusivity, $D$, depends on $Z$. Previous work has shown that for large enough $Z$ changes in diffusivity will approximately follow a power law \cite{roosen2011protein}, and so we can consider $D\left(Z\right)=D_{0}Z^{\gamma}$ which yields
\begin{equation}
Z<\left(\frac{6 D_{0}(A_{0}K_{eq} - P_{0})}{A_{0} k r_{c}^2(1+K_{eq})}\right)^{\frac{1}{1-\gamma}}
\end{equation}
which for $P\approx0$ is
\begin{equation}
Z<\left(\frac{6 D_{0}K_{eq}}{ k r_{c}^2(1+K_{eq})}\right)^{\frac{1}{1-\gamma}}.
\label{z-limit}
\end{equation}
This equation states how the concentration of enzyme should either concentrate or dilute with cell size to avoid thermodynamic limitations. 

Given that $\gamma\approx -3/2$ \cite{roosen2011protein} we have that $Z\propto r_{c}^{-4/5}$ or
\begin{equation}
Z\propto V_{c}^{-4/15}.
\end{equation}
It should be noted that given Equation \ref{z-limit}, $Z$ is bounded by a concentration that is diluting out as cells increase in size. This should set a hard upper bound where the cellular concentrations start to reach discrete numbers of enzymes. Thus, under a fixed tradeoff for diffusivity an upper bound on cell volume is predicted by the dilution of enzyme concentrations required to avoid thermodynamic shutdown. 

It should be noted that once $Z$ reaches sufficiently low concentrations diffusivity saturates to a maximum value set by molecular motion in the fluid \cite{roosen2011protein}. At this point $D\left(Z\right)=D_{max}$ and we have
\begin{equation}
Z <\frac{6 D_{max} A_{0}K_{eq}}{A_{0} k r_{c}^2(1+K_{eq})}.
\end{equation}
In this limit $Z$ will scale like $V_{c}^{-2/3}$ to avoid thermodynamic shutdown.

\subsection*{Thermodynamic Optimization}

\begin{figure}[h!]
    \centering
    \includegraphics[width=0.75\textwidth]{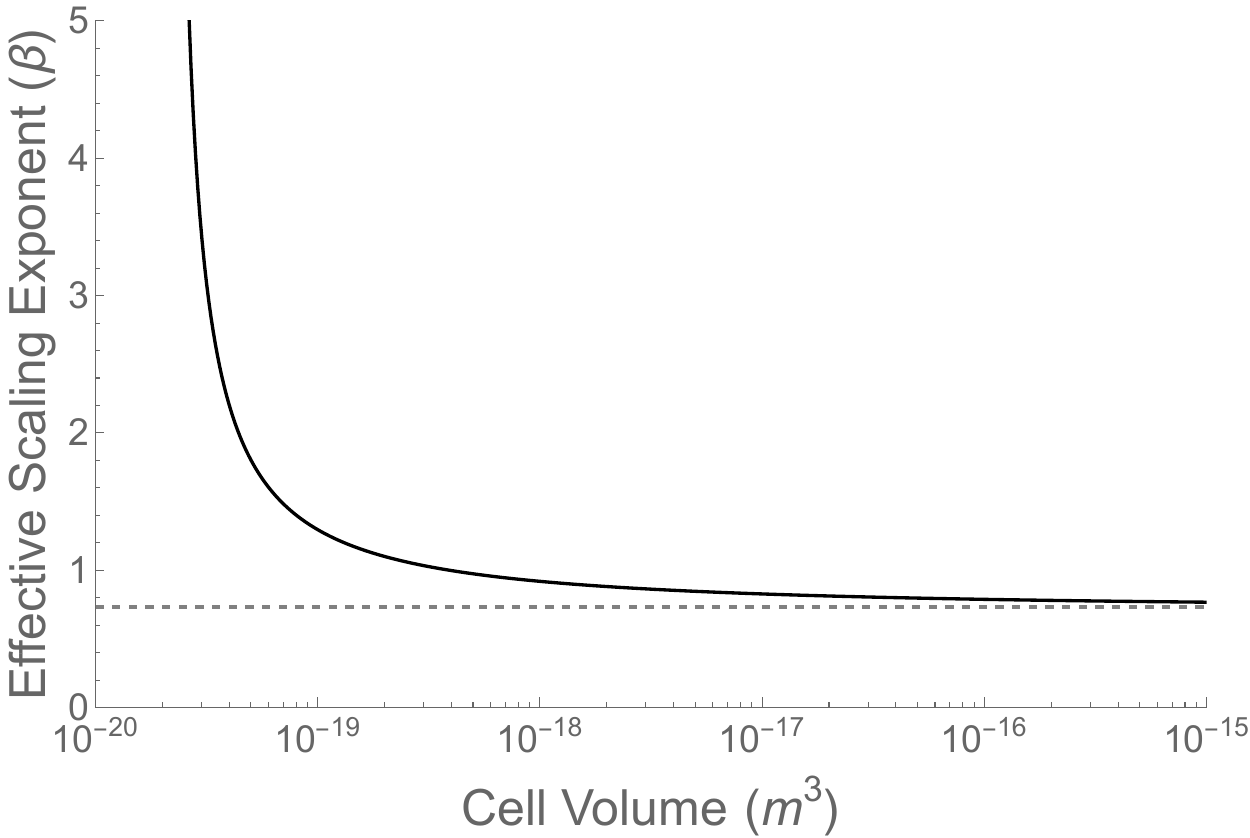}
    \caption{The approximation of the local scaling exponent across the range of bacterial volumes. The dashed line is $11/15$ which the complete model converges on for large cell volumes.}
    \label{SIFig:aproximate-exponent}
\end{figure}

\begin{figure}[h!]
    \centering
    \includegraphics[width=0.75\textwidth]{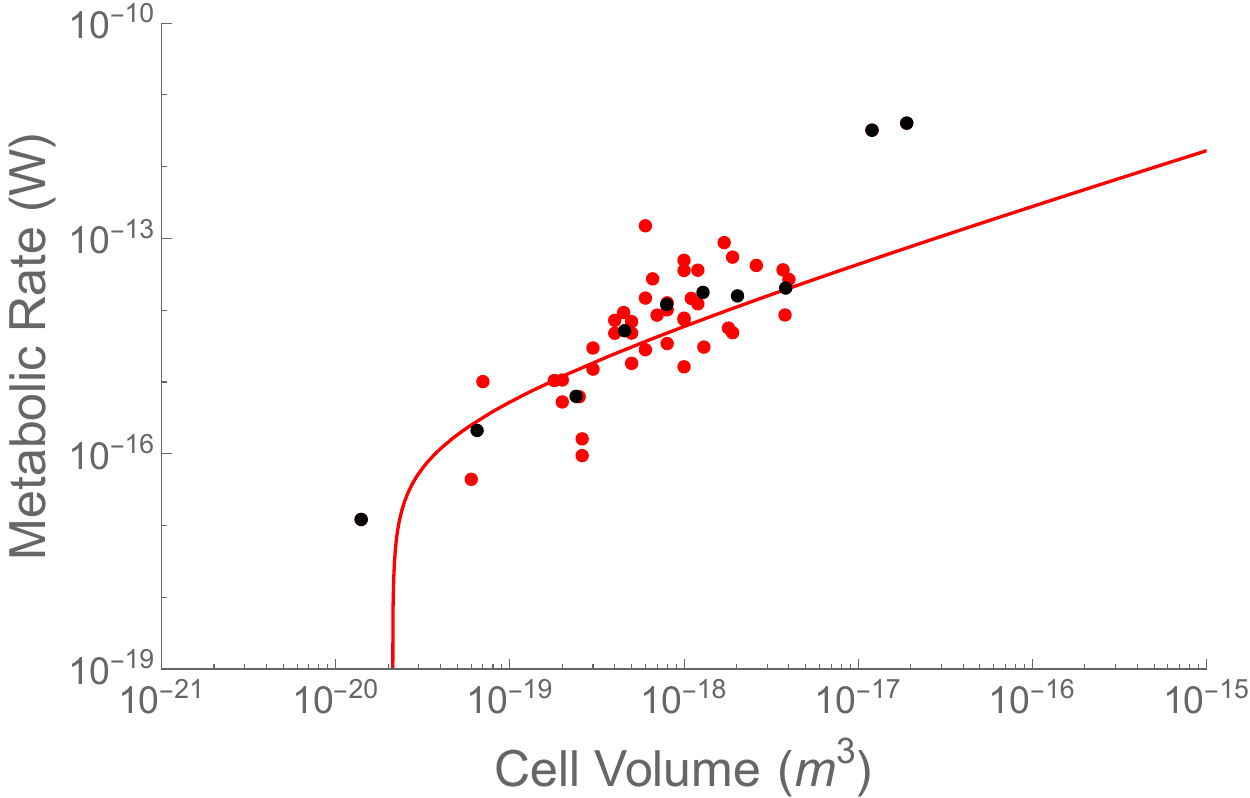}
    \caption{Comparison of the model prediction (red curve) with bacterial data (red points) and the binned averages of those data (black points) same data.}
    \label{SIFig:binned-comparison}
\end{figure}

More generally, we are interested in the concentrations of enzymes that would optimize cellular metabolic rate. The above arguments give us a bounding expectation for the scaling of enzyme concentration, but we can replace these with the complete optimization of cellular metabolic rate, $B$. Specifically, we are interested in 
\begin{equation}
\frac{\partial B}{\partial Z}= 0.
\end{equation}
where the total metabolism integrated over a cell is
\begin{equation}
B= 4 \pi RT k Z \left(1-\frac{3 V_{e}\left(r_{c}\right)}{4\pi  r_{c}^3}\right)
    \int_0^{r_{c}} r^2 A(r) \log \left(\frac{K_{eq} A(r)}{P(r)}\right) dr.
\end{equation}
Considering cases where the environmental concentration of the product is low, $P_{0}\approx 0$, the integrand is given by
\begin{equation}
A_{0} r_{c} r Z \text{csch}\left(\frac{\sqrt{k Z} r_{c}
   }{\sqrt{D(Z)}}\right) \sinh \left(\frac{\sqrt{k Z} r
   }{\sqrt{D(Z)}}\right) \log \left(\frac{K_{eq} r_{c}}{r \sinh
   \left(\frac{\sqrt{k Z} r_{c}}{\sqrt{D(Z)}}\right)
   \text{csch}\left(\frac{\sqrt{k Z} r }{\sqrt{D(Z)}}\right)-r_{c}}\right)
\end{equation}
We approximate this integral considering small $r_{c}$ and with, $D\left(Z\right) = D_{0} Z^{\gamma}$, the dependence of diffusivity on $Z$ . Given these assumptions integrand is well approximated by
\begin{equation}
\frac{A_{0} \sqrt{k} r_{c} r^2 z^{3/2} \text{csch}\left(\frac{\sqrt{k Z} r_{c}}{\sqrt{D_{0} z^{\gamma }}}\right) \log \left(\frac{K_{eq}
   r_{c}}{\frac{\sqrt{D_{0} Z^{\gamma }} \sinh \left(\frac{\sqrt{k Z} r_{c}}{\sqrt{D_{0} Z^{\gamma }}}\right)}{\sqrt{k Z}}-r_{c}}\right)}{\sqrt{D_{0} Z^{\gamma }}}
\end{equation}
leading to the solution of the integral as
\begin{equation}
B=\frac{A_{0} RT k^{3/2} r_{c} Z^{3/2} \left(4 \pi  r_{c}^3-3 V_{e}\right)
   \text{csch}\left(\frac{\sqrt{k Z} r_{c} }{\sqrt{D_{0} Z^{\gamma
   }}}\right) \log \left(\frac{K_{eq} r_{c}}{\frac{\sqrt{D_{0} Z^{\gamma }}
   \sinh \left(\frac{\sqrt{k Z} r_{c} }{\sqrt{D_{0} Z^{\gamma
   }}}\right)}{\sqrt{k Z}}-r_{c}}\right)}{3 \sqrt{D_{0} Z^{\gamma }}}.
\end{equation}
In the small radius limit the derivative, $\partial B/\partial Z$, is then well approximated by
\begin{equation}
\frac{\partial B}{\partial Z} \approx  A_{0} RT k V_{e} \left(1-\gamma -\log \left(\frac{6 K_{eq} D_{0} z^{\gamma
   -1}}{k r_{c}^2}\right)\right).
   \end{equation}
  Setting this to zero leads to 
  \begin{equation}
  Z_{opt}=\frac{1}{e}\left(\frac{6 K_{eq} D_{0}}{k r_{c}^{2}}\right)^{1/\left(1-\gamma\right)}.
  \end{equation}
 This optimal solution of $Z_{opt}$ can be substituted  back into the approximate solution for total metabolic rate, $B$, which we find does a good job of approximating the full numerical integral. If we take $\gamma= -3/2$ this gives 
  \begin{equation}
  Z_{opt}=
  \frac{6^{2/5} K_{eq}^{2/5} D_{0}^{2/5}}{e k^{2/5}
   r_{c}^{4/5}}.
     \end{equation}
This solution shows that the optimal $Z$ scales like 
\begin{equation}
Z\propto r_{c}^{-4/5}\propto V_{c}^{-4/15}
\end{equation}
in agreement with our considerations above for thermodynamic shutdown. The corresponding solution for metabolic rate is given by 
\begin{equation}
B= \frac{2^{9/10} A_{0} RT k^{3/5} K_{eq}^{9/10} D_{0}^{2/5}
   \text{csch}\left(\frac{\sqrt{6} \sqrt{K_{eq}}}{e^{5/4}}\right) \left(4 \pi 
   r_{c}^3-3 V_{e}\right) \log \left(\frac{K_{eq}^{3/2}}{\frac{e^{5/4} \sinh
   \left(\frac{\sqrt{6}
   \sqrt{K_{eq}}}{e^{5/4}}\right)}{\sqrt{6}}-\sqrt{K_{eq}}}\right)}{\sqrt[10]{3}
   e^{9/4} r_{c}^{4/5}}
\end{equation}
which scales like 
\begin{equation}
B\approx r_{c}^{11/5}\propto V_{c}^{11/15}
\end{equation}
predicting that metabolic rate will scale slightly sublinearly for sufficiently large size. It should be note that this is the approximate scaling of $B$ over a range of larger cell sizes, but the full form of $B$ is more complicated a small cells sizes, especially considering the factor of $(1-V_{e}\left(r_{c}\right)/V_{c})$ multiplying the entire scaling where $V_{e}= v_{0} V_{c}^{\alpha}$ with $\alpha=0.83$ \cite{kempes2016evolutionary}. Figure \ref{SIFig:aproximate-exponent} shows the full solution of $B$ with more strongly superlinear scaling because of the small-size curvature.

\subsection*{Saturated Diffusivity}

It should be noted that at very low $Z$ the diffusivity saturates to a constant maximum behavior and stops following a power law. In this case diffusivity reaches the maximum molecular diffusivity, $D=D_{max}$, and the above equations lead a optimal enzyme concentration of 
\begin{equation}
Z_{opt}= \frac{6 D_{max} K_{eq}}{e k r_{c}^2}
\end{equation}
and corresponding metabolic rate of 
\begin{equation}
B= \frac{2 \sqrt{6} A_{0} RT D_{max} K_{eq}^{3/2}
   \text{csch}\left(\sqrt{\frac{6}{e}} \sqrt{K_{eq}}\right) \left(4 \pi 
   r_{c}^3-3 V_{e}\right) \log \left(\frac{1}{\frac{\sqrt{\frac{e}{6}} \sinh
   \left(\sqrt{\frac{6}{e}}
   \sqrt{K_{eq}}\right)}{K_{eq}^{3/2}}-\frac{1}{K_{eq}}}\right)}{e^{3/2}
   r_{c}^2}.
\end{equation}
Here enzyme concentration will scale like $Z\propto V_{c}^{-2/3}$ and $B\propto V_{c}^{1/3}$. However, these scalings are unlikely to be observed because of the shift to employing active transport, and these results are only shown to illustrate the extreme limitation faced by large cells if they only used molecular diffusion.

\section{Active Transport Tradeoffs} 
Our model for active transport treats diffusion in the cell as
\begin{eqnarray}
\frac{\partial A}{\partial t}&=&D^{\prime}\frac{1}{r^{2}}\frac{\partial}{\partial r}\left(r^{2}\frac{\partial A}{\partial r}\right)-k_{cat}A Z   \nonumber \\
\frac{\partial P}{\partial t}&=&D^{\prime}\frac{1}{r^{2}}\frac{\partial}{\partial r}\left(r^{2}\frac{\partial P}{\partial r}\right)+k_{cat}A Z_0 
\end{eqnarray}
with a modified effective diffusivity, $D^{\prime}$, that follows
\begin{equation}
D^{\prime}=(1-\rho)D_{max}+\rho D_{trans}.
\end{equation}
where $\rho$ is the fraction of cell volume occupied by transport structures. Here we define an effective volume (region with enhanced transport) of a transport structure, and $\rho$ is the ratio of the total of these effective volumes to the total cell volume. The optimization of metabolism now depends on tradeoffs between the adjusted diffusivity and the space occupied by the transport structures, where total metabolism follows
\begin{equation}
B= 4 \pi RT k Z \left(1-\rho \epsilon\right)
    \int_0^{r_{c}} r^2 A(r) \log \left(\frac{K_{eq} A(r)}{P(r)}\right) dr.
\end{equation}
where $\epsilon$ is the ratio of the volume of a transport structure to the effective volume of enhanced transport that structure creates. The solutions for $A\left(r\right)$ and $P\left(r\right)$ are the same as in Equations \ref{Ar-solution} and \ref{Pr-solution}.

We are again interested in optimizing cellular metabolic rate, $B$, where now we are looking for solutions to
\begin{equation}
\frac{\partial B}{\partial \rho}= 0.
\end{equation}
The integration of $B$ is well approximated by
\begin{equation}
B\approx \frac{C (1-\rho  \epsilon)
   \text{csch}\left(\frac{\sqrt{k Z} r_{c} }{\sqrt{\rho D_{trans}  +D_{max}
   (1-\rho)}}\right) \log \left(\frac{K_{eq}
   r_{c} \sqrt{k Z}}{\sqrt{\rho D_{trans} +D_{max} (1-\rho)} \sinh
   \left(\frac{\sqrt{k} r_{c} \sqrt{Z}}{\sqrt{\rho D_{trans} +D_{max} (1-\rho
   )}}\right)-r_{c}}\right)}{3 \sqrt{\rho D_{trans}
   +D_{max} (1-\rho)}}.
 \end{equation}
 \begin{equation}
 C\equiv 4 \pi  A_{0} RT k^{3/2} r_{c}^4 Z^{3/2}
 \end{equation}
 for small values of $r_{c}$. Again we are considering the regime where transport, now represented by the adjusted diffusivity $D^{\prime}$, is large relative to the scale of the cell. Taking the derivative of this this solution and again approximating it for small $r_{c}$ and also for small $D_{max}$ (the maximum molecular diffusivity) we have 
 \begin{equation}
\frac{\partial B}{\partial \rho}\approx \frac{r_{c}^3 A_{0} RT k Z \left(1 - \rho \epsilon \left(1+ \log \left(\frac{6 D_{trans}
   K_{eq} \rho }{k r_{c}^2 Z}\right)\right) \right)}{3 \rho } =0 .
 \end{equation}
which gives the optimal fraction of transport structures as
\begin{equation}
\rho = \frac{1}{\epsilon  W\left(\frac{6 e D_{trans} K_{eq}}{k r_{c}^2 Z \epsilon
   }\right)}
\end{equation}
where $W$ is the Lambert $W$ function (the product logarithm). For small $r_{c}$ this is well approximated by 
\begin{equation}
\rho = \frac{1}{\epsilon  \log \left(\frac{6 e D_{trans} K_{eq}}{k r_{c}^2 Z \epsilon
   }\right)}
\end{equation}
Similar to our solutions above for the enzyme concentration tradeoffs in bacteria, this function again highlights that optimal metabolic values depend on $\frac{D K_{eq}}{k r_{c}^{2}}$.

\begin{figure}[h!]
    \centering
    \includegraphics[width=0.75\textwidth]{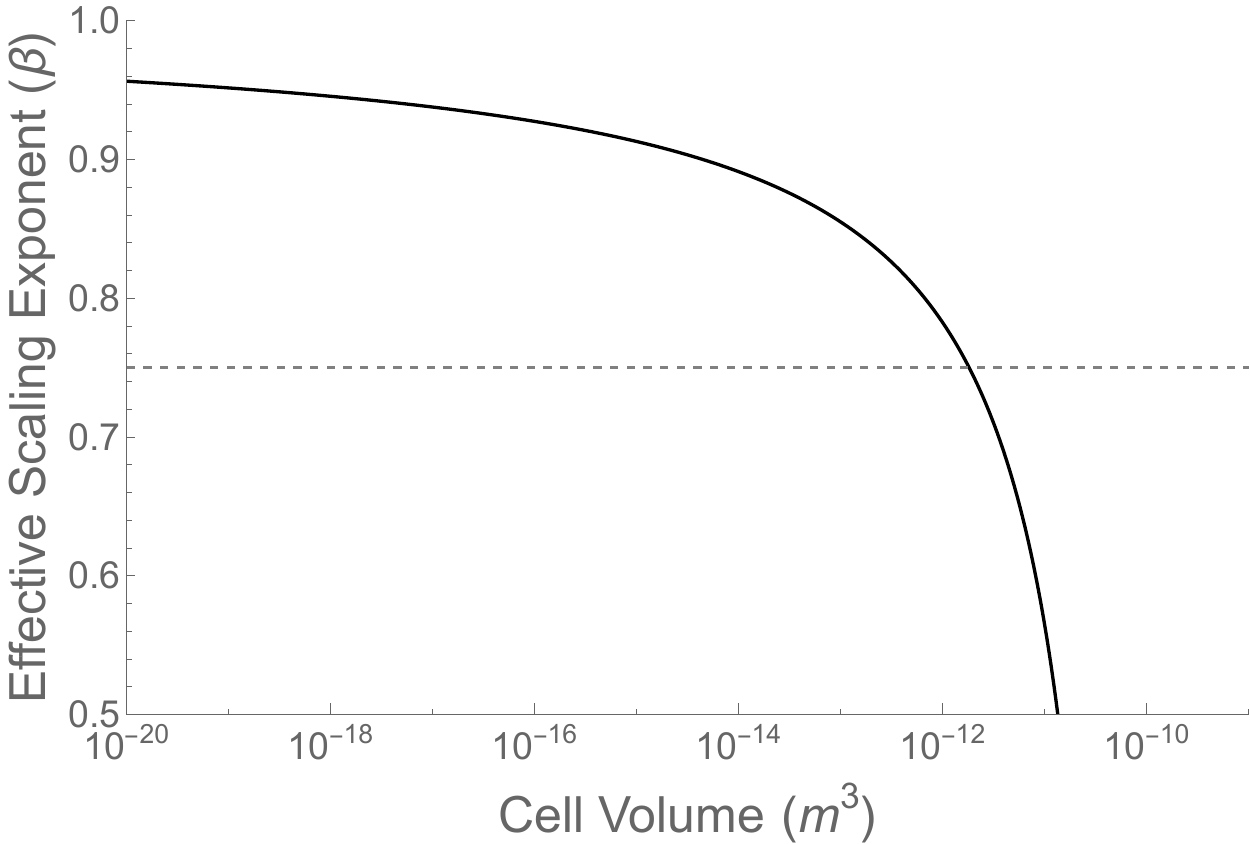}
    \caption{The approximation of the local scaling exponent across the range of bacterial volumes. The dashed line is $11/15$ which the complete model converges on for large cell volumes.}
    \label{SIFig:euk-aproximate-exponent}
\end{figure}

Taking all of this together we have that the optimal metabolic rate follows
\begin{equation}
B=\frac{1}{3}A_{0} RT k r_{c}^3 Z \left(1+1/\log \left(\frac{6 D_{trans} K_{eq}}{k r_{c}^2
   Z \epsilon }\right)\right) \log \left(\frac{6 D_{trans} K_{eq}}{k r_{c}^2 Z \epsilon \left(1+
    \log \left(\frac{6 D_{trans} K_{eq}}{k r_{c}^2 Z \epsilon }\right)\right) }\right).
    \end{equation}
 This scales like $B\propto r_{c}^{3} \propto V_{c}$ with a logarithmic correction term. This explains why the scaling of metabolism in Eukaryotes is close to linear with cell volume but actually sublinear due to the logarithmic terms. Figure \ref{SIFig:euk-aproximate-exponent} gives the approximate local exponent of $B$ with cell size. The full range of Eukaryotes is well approximated by $\beta=0.90$ which agrees with data as discussed in the maintext. 
 
 This function also comes with an asymptotic limit that occurs because transportors begin to overpack the cell. This limit occurs when 
 \begin{equation}
 W\left(\frac{6 e D_{trans} K_{eq}}{k r_{c}^2 Z \epsilon }\right)=1
 \end{equation}
  which corresponds to an maximum single cell eukaryote of
 \begin{equation}
 V_{c,max}=\frac{8 \sqrt{6} \pi  D_{trans}^{3/2} K_{eq}^{3/2}}{k^{3/2} Z^{3/2} \epsilon ^{3/2}}.
  \end{equation}
 This solution mirrors the molecular diffusion limit but with an altered diffusivity.

\end{document}